\documentclass[sigconf,authordraft,review=false]{acmart}

\usepackage{amsmath,amssymb,amsfonts}
\usepackage{graphicx}
\usepackage{textcomp}
\usepackage{xcolor}
\usepackage{algorithm}
\usepackage{algpseudocode}
\usepackage{listings}
\usepackage{color}
\usepackage{seqsplit}

\definecolor{dkgreen}{rgb}{0,0.6,0}
\definecolor{gray}{rgb}{0.5,0.5,0.5}
\definecolor{mauve}{rgb}{0.58,0,0.82}

\usepackage{url}

\definecolor{background}{HTML}{EEEEEE}

\colorlet{punct}{red!60!black}
\definecolor{background}{HTML}{EEEEEE}
\definecolor{delim}{RGB}{20,105,176}
\colorlet{numb}{magenta!60!black}

\lstdefinelanguage{json}{
    comment=[l]{//},
    morecomment=[s]{/*}{*/},
    literate=
      *{:}{{{\color{punct}{:}}}}{1}
      {,}{{{\color{punct}{,}}}}{1}
      {\{}{{{\color{delim}{\{}}}}{1}
      {\}}{{{\color{delim}{\}}}}}{1}
      {[}{{{\color{delim}{[}}}}{1}
      {]}{{{\color{delim}{]}}}}{1},
}

\AtBeginDocument{%
  \providecommand\BibTeX{{%
    \normalfont B\kern-0.5em{\scshape i\kern-0.25em b}\kern-0.8em\TeX}}}

\setcopyright{acmcopyright}
\copyrightyear{2020}
\acmYear{2020}
\acmDOI{10.1145/1122445.1122456}

\acmConference[MSR '20]{17th International Conference on Mining Software Repositories}{May 25--26, 2020}{Yongsan-gu, Seoul, South Korea}
\acmBooktitle{MSR '20: 17th International Conference on Mining Software Repositories, May 25--26, 2020, Yongsan-gu, Seoul, South Korea}

\acmPrice{15.00}
\acmISBN{978-1-4503-XXXX-X/18/06}

\begin{document}

\title[Does Unit-Tested Code Crash?]{Does Unit-Tested Code Crash? \\ A Case Study of Eclipse}

\author{Efstathia Chioteli}
\authornote{Chioteli collected the unit testing results and performed
the qualitative analysis.
Batas performed the quantitative analysis.
All authors contributed equally to the paper's writing.}
\email{t8150148@aueb.gr}

\author{Ioannis Batas}
\authornotemark[1]
\email{t8150090@aueb.gr}

\author{Diomidis Spinellis}
\authornotemark[1]
\email{dds@aueb.gr}
\orcid{0000-0003-4231-1897}

\affiliation{%
 \institution{Department of Management Science and Technology\\
                    Athens University of Economics and Business}
  \city{Athens}
  \country{Greece}
}

\renewcommand{\shortauthors}{Chioteli, Batas, Spinellis}

\newcommand{\+}{\discretionary{\mbox{$\hookleftarrow$}}{}{}}
\renewcommand{\seqinsert}{\ifmmode\allowbreak\else\+\fi}
\newcommand{\li}[1]{\texttt{\seqsplit{#1}}}

\begin{abstract}
\textit{Context:}
Software development projects increasingly adopt unit testing as a way
to identify and correct program faults early in the construction process.
Code that is unit tested should therefore have fewer failures associated
with it.
\\
\textit{Objective:}
Compare the number of field failures arising in unit tested code against
those arising in code that has not been unit tested.
\\
\textit{Method:}
We retrieved 2\,083\,979 crash incident reports associated with the Eclipse
integrated
development environment project,
and processed them to obtain a set of 126\,026 unique program failure
stack traces associated with a specific popular release.
We then run the JaCoCo code test coverage analysis on the same release,
obtaining results on the line, instruction, and branch-level
coverage of 216\,392 methods.
We also extracted from the source code 
the classes that are linked to a corresponding test class so as to 
limit test code coverage results to 1\,267 classes with actual tests.
Finally, we correlated unit tests with failures
at the level of 9\,523 failing tested methods.
\\
\textit{Results:}
Unit-tested code does not appear to be associated with fewer failures.
\\
\textit{Conclusion:}
Unit testing on its own may not be a sufficient method for preventing
program failures.
\end{abstract}



\maketitle

\section{Introduction} 
The rising size and complexity of software multiply the demands
put on adequate software testing~\cite{KTL15}.
Consequently,
software development projects increasingly adopt unit testing~\cite{BG98}
or even test-driven development~\cite{Bec03} as a way
to identify and correct program faults early in the construction process.
However, the development of testing code does not come for free.
Researchers have identified that one of the key reasons for the limited
adoption of test-driven development is the increased development time~\cite{CSP11}.
It is therefore natural to wonder whether the investment in testing
a program's code pays back through fewer faults or failures.

One can investigate the relationship between the software's production code
and its tests by utilizing heuristics or code coverage analysis~\cite{WMMK01}.
Heuristics are based on conventions associated with the development
of unit test code;
for example that a test class is named after the class it tests
(e.g. {\em Employer}),
followed by the {\em Test} suffix (i.e. {\em EmployerTest}).
Code coverage analysis is a process that provides
metrics indicating to what extent code
has been executed---under various control flow measures~\cite{BMS98}.
The corresponding metrics can be efficiently obtained through diverse
tools~\cite{TH02,Hof11}.
Then, the process for determining test coverage
involves running the software's test suite,
and obtaining code coverage metrics,
which in this case indicate code that (probably) is
or (definitely) is not tested.

To examine how test code coverage relates to software quality,
numerous methods can be employed.
One can look at corrected faults and see whether the corresponding
code was tested or not~\cite{MND09,KLLN17}.
In addition, faults can be deliberately introduced by mutating the
code~\cite{JH11} in order to look at how test code coverage relates
to test suite effectiveness~\cite{GJG14,IH14}.
Alternatively, one can artificially vary test coverage to see its
effect on exposing known faults~\cite{KTL15}.
Finally, one could look at software failures rather than faults
and correlate these with test code coverage.

In this study we investigate the relationship between unit testing and
failures by examining the usage of unit testing on code that is
associated with failures in the field.
We do this in three conceptual steps.
First, we run software tests under code coverage analysis
to determine which methods have been unit tested and to what extent.
We triangulate these results with heuristics regarding the naming of
classes for which unit test code actually exists.
Then, we analyze the stack traces associated with software failure
reports to determine which methods were associated with a specific
failure.
Finally, we combine the two result sets and analyze
how unit-tested methods relate to observed failures.

We frame our investigation in this context
through the following research questions.\\
\textbf{RQ1} How does the testing of methods relate to observed failures?\\
\textbf{RQ2} Why do unit-tested methods fail?

A finding of fewer failures associated with tested code would support
the theory that unit testing is effective in improving software reliability.
Failing to see such a relationship would mean that further research is required
in the areas of
unit test effectiveness (why were specific faults not caught by unit tests) and
test coverage analysis (how can coverage criteria be improved to
expose untested faults).

The main contributions of our study are the following:
\begin{itemize}
\item a method for investigating the effectiveness of unit testing,
\item an empirical evaluation between unit test coverage and failure reports, and
\item an open science data set and replication package
providing empirical backing and replicability for our findings.
\end{itemize}

In the following sections we
describe the methods we used (Section~\ref{sec:methods}),
present our quantitative and qualitative results (Section~\ref{sec:results}),
discuss their implications (Section~\ref{sec:discussion}),
examine the threats to the validity of our findings (Section~\ref{sec:threats}),
outline related work in this area (Section~\ref{sec:related}), and
conclude with a summary of our findings and their implications
(Section~\ref{sec:conclusions}).

\section{Methods} 
\label{sec:methods}
We based our study on the popular Eclipse
open source integrated development environment~\cite{Gee05}.
In brief,
to answer our research questions
we obtained data regarding failures of the Eclipse IDE,
we determined the most popular software version associated with the failures,
we built this specific software version,
we run the provided tests under a code coverage analysis tool,
we combined the results with heuristics regarding the naming of test code,
we joined the analyzed software failures with the corresponding code coverage analysis results,
and we analyzed the results through statistics and a qualitative study.
Following published recommendations~\cite{Ince2012},
the code and data associated with this endeavor (AERI JSON data,
code coverage analysis, stack traces analyzed,
analysis scripts,
and combined results) are openly available
online.\footnote{\url{10.5281/zenodo.3610822}}

In our presentation we use the following terms as defined in the
systems and software engineering vocabulary standard~\cite{ISO24765-2017}.
\begin{description}
\item[Error] ``Human action that produces an incorrect result.''
\item[Fault] ``Incorrect step, process, or data definition in a
computer program'';
``defect in a system or a representation of a system that if
executed/activated could potentially result in an error.''
\item[Failure]
``Termination of the ability of a system to perform a required function
or its inability to perform within previously specified limits;
an externally visible deviation from the system's specification.''
\end{description}
According to these definitions,
a programmer {\em error} may result in a {\em fault} in the code.
This may in turn cause a {\em failure} in the program's operation,
which may manifest itself as e.g. incorrect output,
a program freeze, or an abnormal program termination;
the last one often accompanied by a diagnostic report,
such as a stack trace.

\subsection{Data Provenance and Overview} 
We conducted our research on a dataset of anonymized diagnostic
failure reports communicated to the Eclipse developers
through the IDE's Automated Error Reporting (AERI)
system~\cite{Bal18}.
The AERI system is installed by default on the Eclipse IDE
to aid support and bug resolution.
Its back-end collects {\em incidents},
which contain data regarding a particular instance of an uncaught exception.
It analyses them and aggregates similar ones into {\em problems}.
The specific data set we used was generated on 2018-02-17 and contains data
collected over the period 2016-03-13 to 2016-12-13.
The dataset contains a file with the complete collected data and
two extracts in CSV format containing a subset of fields
and aggregate data.
We based our study on the set titled ``All Incidents'',
which consists of 2\,083\,979 crash reports provided in the form of JSON
files.\footnote{\url{http://software-data.org/datasets/aeri-stacktraces/downloads/incidents_full.tar.bz2}}

\begin{lstlisting}[language=json,
float,
caption={AERI incident data extract},
label={l:aeri}
]
// incident_1040416.json

{
  "eclipseBuildId": "4.5.2.M20160212-1500",
  "eclipseProduct": "org.eclipse.epp.package.java.product",
  "fingerprint": "b8623a6f9da69438eae9e21911c9e8ca",
  "fingerprint2": "bbf2fcfe645cea0dc60d3d521d530b84",
  "javaRuntimeVersion": "1.8.0_91-b14",
  "kind": "NORMAL",
  "osgiArch": "x86_64",
  "osgiOs": "Windows7",
  "osgiOsVersion": "6.1.0",
  "osgiWs": "win32",
  "presentBundles": [
    { "name": "org.eclipse.core.commands", "version": "3.7.0.v20150422-0725" },
    ......
  	],
  "savedOn": "2016-07-12T14:00:32.468Z",
  "severity": "UNKNOWN",
  "stacktraces": [
    [
    	  // Topmost frame; part of the top-1/6/10 frames
      {
        "cN": "org.eclipse.jdt.internal.ui.JavaPlugin",
        "fN": "JavaPlugin.java",
        "lN": 320,
        "mN": "log"
      }, /* [...] Four more frames here */ 
      // Frame 6; part of the top-6 and top-10 frames
      {
        "cN": "org.eclipse.jface.text.contentassist. ContentAssistant",
        "fN": "ContentAssistant.java",
        "lN": 1902,
        "mN": "computeCompletionProposals"
      },  /* [...] Three more frames here */ 
      // Frame 10; part of the top-10 frames
      {
        "cN": "org.eclipse.swt.custom.BusyIndicator",
        "fN": "BusyIndicator.java",
        "lN": 70,
        "mN": "showWhile"
      }, ... // More stack frames; not examined
    ]
   ],
  "status": {
    "code": 4,
    "fingerprint": "2a4caf19f4a424c54ea1951d14ec3341",
    "message": "An error occurred while computing quick fixes. Check log for details.",
    "pluginId": "org.eclipse.jdt.ui",
    "pluginVersion": "3.11.2.v20151123-1510",
    "severity": 4
  },
  "timestamp": "2016-07-12T14:00:32.430Z"
}  
\end{lstlisting}

As subset of an AERI report in JSON format appears in
Listing~\ref{l:aeri}.
The data attributes that are interesting for
the purpose of our research are the following:
\begin{itemize}
\item \textit{EclipseProduct}, the product associated with the Eclipse project,
\item \textit{BuildId}, the version of the Eclipse source code, and
\item \textit{Stacktrace}, the incident's stack trace. Each stack trace
consists of successive stack frames and their details (class, method, line).
\end{itemize}

To combine incident reports with the associated source code,
compiled code, and test data, we decided to focus our study on a
specific version of Eclipse.
We therefore analyzed the AERI incident reports to find
the Eclipse release associated with the highest number.
This would allow us to obtain a large dataset for statistical analysis.
Given that production releases are widely distributed,
numerous failures associated with a release are a sign of the release's
popularity, rather than its inherent instability.
An overview of the incidents associated with each release can be found in
the AERI incidents analysis report~\cite[p. 18]{Bal18}.
The data corresponding to the selected version
consist of 126\,026 incident files that have
{\em EclipseProduct} equal to \li{org.eclipse.epp.package.java.product}
and
{\em BuildID} equal to \li{4.5.2.M20160212-1500}.

\subsection{Generation of Test Code Coverage Data} 
\label{sec:jacoco}
To create code coverage data associated with tests
we needed to obtain the source code,
compile it, and perform code coverage analysis while running the tests.

We accessed the Eclipse source code through the Eclipse Platform {\em Releng}
project,\footnote{\url{https://wiki.eclipse.org/Platform-releng/Platform_Build}}
which provides instructions for building the Eclipse Platform using
preferred technologies identified as part of the
Eclipse Common Build Infrastructure (CBI) initiative.
This combines infrastructure, technologies, and practices
for building Eclipse software.
To ensure that test coverage results would be coeval with the corresponding
incident reports, we retrieved the source code version of Eclipse corresponding
to the one whose stack traces we chose to analyze.

To obtain data regarding Eclipse's test coverage, we used the
JaCoCo Code Coverage~\cite{Hof11} system,
which is an open-source toolkit for measuring and reporting Java code
coverage.  
It reports instruction, branch, line, method, class, package, and complexity coverage.  
Instruction is the smallest unit JaCoCo counts and is associated with 
single Java byte code instructions.
Branch coverage reports on taken and non-taken branches.
Cyclomatic complexity coverage reports the ratio of
the executed cyclomatic complexity~\cite{McC76} graph paths over the
total cyclomatic complexity number.
Under line coverage, a source code line is considered to be covered if
at least one instruction associated with the line has been executed.
Finally, coverage of larger aggregates is reported on the basis of
at executing at least a single instruction.
For the purposes of this study we focused on the most fine grained
coverage metrics, namely line, instruction, and branch coverage.

Eclipse is a multi-module project,
which hinders the derivation of code coverage reports,
because the JaCoCo Maven goals used to work on single modules only.
For that reason, we used the new ``Maven Multi-Module Build''
feature,\footnote{\url{https://github.com/jacoco/jacoco/wiki/MavenMultiModule}} 
which implements a Maven goal called ``jacoco:report-aggregate''.
This aggregates coverage data across Maven modules.

To apply this feature, 
we first added the JaCoCo plugin and profile in the Maven parent \li{pom.xml}
file, and then we created a separate project where we:
\begin{itemize}
\item configured the \li{report-aggregate} goal,
\item added as dependencies with \li{scope compile} the projects
containing the actual code and with \li{scope test} the projects
containing the tests and the \li{.exec}-suffixed data.
\end{itemize}

Owing to the size and complexity of Eclipse,
the process of compiling, testing, and obtaining code coverage
results was far from trivial.

First, due to the fact that we run JaCoCo on an old (by about three years)
release of Eclipse, it was difficult to make the code compile.
The release comprised some repositories which had become archived at
the time we attempted to build it, and therefore the specified
path could not found.
For example, in one case Maven terminated with an internal error
reporting that it had failed to load to the repository
\li{eclipse-p2-repo} from location
\li{http://download.eclipse.org/eclipse/updates/4.5-M-builds}.
To address this issue we replaced these repositories with newer ones.

Second, three tests remained stuck for more than one hour.
One reason we might think this was happening was because the test was
repeatedly trying to find a specific file.
In the end we had to manually remove the offending tests in
order to proceed with testing and code coverage analysis process.

Third, there were tests that needed specific 
configurations and data to run.
Again, we had to skip those tests  in order to allow the remaining
ones to run.

These three problems resulted in a very time-consuming process,
because every time a run failed, we had to fix and restart it
to find the next missing repository or stuck test.
The average time of each run was about three hours.
In total we spent around three months fine-tuning the compilation
process and the tests, until we were able to compile the project from
source and run most of the tests to obtain test code coverage results.

\subsection{Determination of Unit Test Classes} 
\label{sec:classes}

Code coverage data can be a notoriously fallacious measure of
test quality~\cite{Sto05}.
Because good quality tests are associated with high code coverage
and the absence of tests with low code coverage,
many mistakenly think that high code coverage implies 
good quality testing.
In fact, high code coverage can be achieved by having some code
executed without testing its correct behavior.
As an example the code for setting up a test case can invoke
some class constructors, from the same or from another class,
without however checking that the corresponding objects are
correctly constructed.

To alleviate false positives regarding the existence of tests
that would result from naively analyzing code coverage reports
(class $A$ got executed, therefore it is tested),
we combined method-level code coverage data with data
regarding the existence of classes containing unit tests.

Specifically, we found the classes of the source code that are relevant to 
a test class with the following procedure. 
A common unit test naming practice is to add the word 'Tests' in the end of the class name. 
However, that is not the case for the test methods (whose name may be quite different) 
and so we focused only on finding the unit-tests on class level.

Through manual examination of the source code we determined that
Eclipse's source code test files (classes) are usually named as follows.
\begin{enumerate}
\item
{\em ClassName}\textbf{Tests}
\item
{\em ClassName}\textbf{Test}
\item
\textbf{Test}{\em ClassName}
\item
\textbf{Test\_PackageName\_}{\em ClassName}
\item
{\em ClassName}\textbf{Tester}
\end{enumerate}

Another common unit test practice is to place test files under the tests/ folder 
with similar path as the class that they test. For example:
\begin{description}
\item[Bundle Class:]
\li{eclipse.platform.ui/\textbf{bundles}/org.eclipse.jface.databinding/src/org/eclipse/jface/databinding/swt/WidgetProperties.java}
\item[Test Class:]
\li{eclipse.platform.ui/\textbf{tests}/org.eclipse.jface.tests.databinding/src/org/eclipse/jface/\textbf{tests}/databinding/swt/WidgetPropertiesTest.java}
\end{description}
In the example above, the path is exactly the same apart from the additional 'tests/' folders. 
However, there were cases on the Eclipse source code,
where the test of the class 
was not placed in the exact path, but we also matched these files since the test class was related 
to the bundle class. For example:
\begin{description}
\item[Bundle Class:] \li{eclipse.platform.ui/bundles/org.eclipse.jface/src/org/eclipse/jface/preference/BooleanFieldEditor.java}
\item[Test Class:] \li{eclipse.platform.ui/tests/org.eclipse.ui.tests/Eclipse JFace Tests/org/eclipse/jface/tests/preferences/BooleanFieldEditorTest.java}
\end{description}

We wrote a script to find the classes that have a corresponding unit test
file by devising through successive experiments
and implementing the following heuristics.
For each sub-module  we generated two sets:
one of files containing in their name the word ``test'' and its
complement.
We then matched the two sets, by removing the word ``test'' from the
filename and also by traversing the associated paths from the right to
the left.
In cases where this method failed, we matched files based on the number
of same words in each path.
In all cases but one we had one or more test files match a single class file.
The cases where this relationship did not hold concerned the separate
implementations of SWT for the Cocoa, GKT, and Win32 back-ends,
which all shared the same test class \li{eclipse.platform.swt/tests/org.eclipse.swt.tests/JUnit Tests/org/eclipse/swt/tests/junit/Test\_org\_eclipse\_swt\_widgets\_Text.java}.
Also in another 15 cases we had the name of multiple test files correspond
to multiple classes; we paired these tests with the corresponding
classes through manual inspection.
Through this procedure we matched 1\,308 test files with 1\,267 classes.

In addition, for each bundle class and each test class we counted the number of lines so as to see 
how well a class is tested (test density).

\subsection{Data Synthesis} 
\label{sec:datasynthesis}

As a next step we combined the data elements derived in the preceding
steps as follows.
\begin{enumerate}
\item
Process the incident files of the dataset, 
extracting all methods from the stack traces together with their
order of appearance. If a method appeared twice in the stack trace
we kept only the very first appearance to avoid duplications.
\item
Process the XML file generated by the JaCoCo coverage report
(Section~\ref{sec:jacoco}),
extracting all methods together with their code coverage data. 
\item
Process the unit test classes file and their line density generated with the 
method explained in Section~\ref{sec:classes}.
\item
Join the common methods of the three preceding lists into
a new list containing the combined fields.
\end{enumerate}
The resulting output and its description are provided in the paper's
replication package.

\subsection{Preliminary Quantitative Analysis} 
\label{sec:quant}
\begin{figure}
{
\includegraphics[width=0.55\columnwidth]{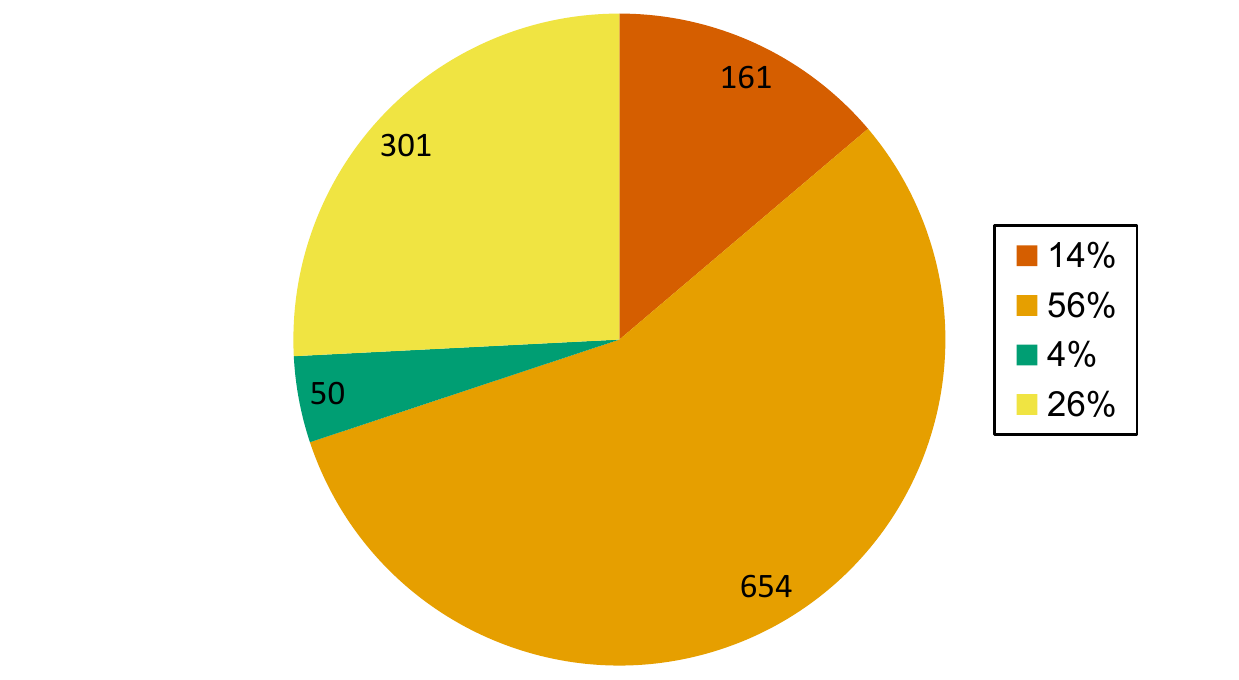}
\includegraphics[width=0.42\columnwidth]{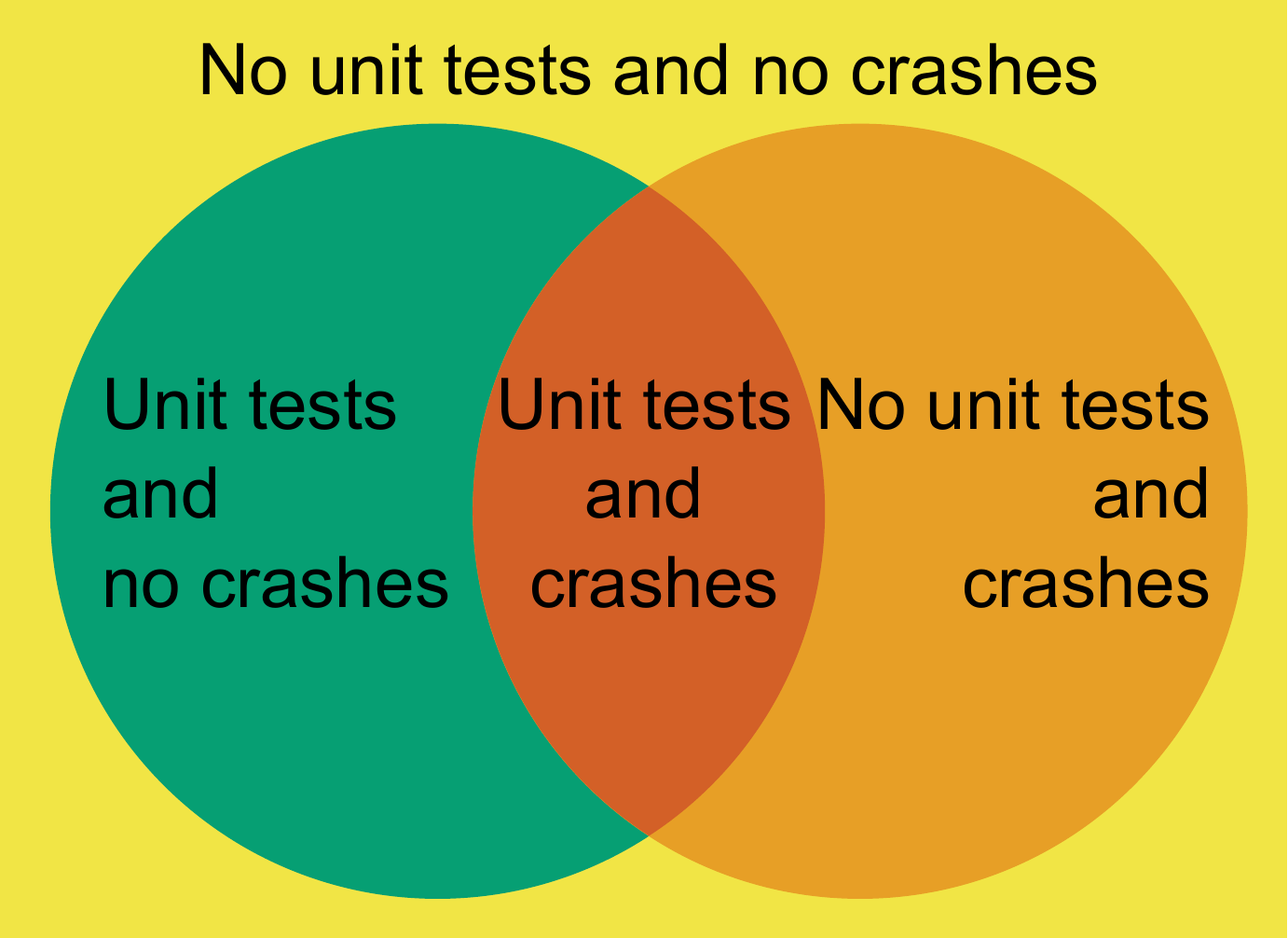}
}
\caption{Relationship between test code coverage at the level methods
and existence of unit tests at the level of classes.
The pie areas correspond to the colored areas of the Venn diagram
depicted on the right.
}
\label{fig:sharepie}
\end{figure}

Overall, JaCoCo tallied 31\,181 Eclipse source code classes
of which the 10\,513 were covered (34\%). 
At the method level,
out of 216\,392 methods the 71\,238 (33\%) were reported as covered.
In addition, JaCoCo  reported that:

\begin{itemize}
\item
1\,960\,447 instructions out of 5\,788\,907 were covered (34\%).
\item
217\,622 branches out of 686\,481 were covered (32\%), and
\item
481\,625 of lines out of 1\,413\,785 were covered (34\%),
\end{itemize}
Based on the small variation of the above figures, we decided
to base our analysis on test code line coverage, which is the
most frequently used method.

\begin{figure}
\centering
\includegraphics[width=0.45\textwidth,trim={2cm 1cm 0cm 1cm}]{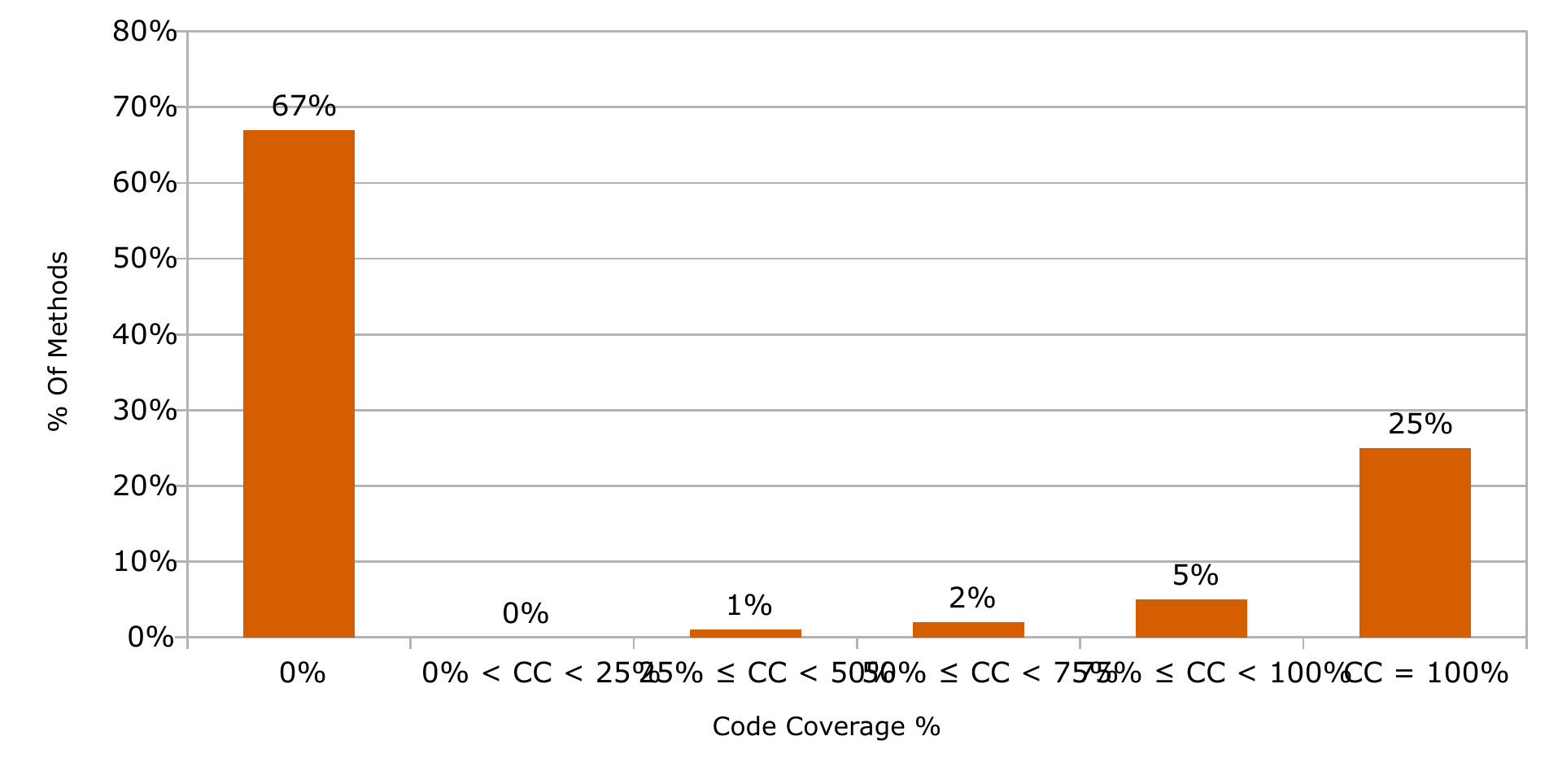}
\caption{Relationship between code coverage and methods of JaCoCo report.}
\label{fig:ccoverall}
\end{figure}

Given the variation of code coverage within a method's body,
we also examined coverage of methods in terms of lines.
Interestingly,
25\% of the methods are fully covered, out the 33\% that are covered in total
(Figure~\ref{fig:ccoverall}).

In parallel with the JaCoCo code coverage, we found that 1\,267 classes 
belong to a class with unit-test code (Section~\ref{sec:classes}).
The small number of unit-tested classes against the total number of
Eclipse's classes can be partly justified by the fact that
abstract classes (about two thousand)
and interfaces (about six thousand)
typically do not have methods to test.
The total lines of these classes 
are 683\,540 and the total lines of the unit-tested classes is 454\,938,
giving a test code line density ratio of 67\% and a median test code line
density across methods of 71\%.

We see that only 12\% (1\,267/10\,513) of the JaCoCo covered classes 
belong to unit-tested classes. This small number is justified by the fact that 
JaCoCo shows which code was executed (or not) when running tests. The
coverage does not necessarily mean that a given class was tested, because its code 
might have been called from other code.

The position of faulty methods in a stack trace was examined by
Schroter and his colleagues~\cite{schroter2010stack}.
They studied 2\,321 bugs from the Eclipse project,
and examined where defects were located in stack traces as
defined by the corresponding fix.
Their research showed that
40\% of bugs were fixed in the very first frame,
80\% of bugs were fixed within the top-6 stack frames, and
90\% of bugs were fixed within the top-10 stack frames.
We correspondingly grouped and matched methods appearing in stack traces
into three groups of methods:
those that have appeared at least once
in the very first,
in the top-6, and
in the top-10 stack frames.

Following the methods we described in Section~\ref{sec:datasynthesis},
we matched 14\,902 crash methods with their test coverage details.
Since faulty methods in a stack trace appear mostly,
within the top-10 stack frames~\cite{schroter2010stack},
we excluded the methods that appeared after the 10th stack frame 
and kept 9\,523 crash methods.

Of those 9\,523 crash methods, the 6\,042 (63\%) are covered according 
to JaCoCo (${\rm instruction\%} > 0$). In terms of branches they are covered 
by 55\%, in terms of instructions by 58\% and in terms of lines, by 59\%.
Also, 1\,518 (16\%) methods out of 9\,523 belong to a class  with unit test, 
with average code coverage line density of 71\%.
The number of methods that belong to a class with unit test and have 
also non-zero code coverage percentage, is 1\,152. 

We decided to focus only on the methods appearing in 
the stack trace's top position, because
a) the method at the stack's top, as the one where the
exception occurred, is certainly implicated
in the crash, even if it may not be the crash's root cause; and
b) 40\% of bugs are fixed in the very first frame~\cite{schroter2010stack}.
The number of crash methods that satisfy this criterion, 
is 1\,166 methods out of the 9\,523 (12\%).

\begin{table}
\caption{Code Coverage of Methods with Class Unit Tests}
\label{tab:ccvstest}
\begin{centering}
\begin{tabular}{lrrrrrr}
\hline
 & \multicolumn{4}{c}{Class Unit Test} & & \\
Covered & \multicolumn{2}{c}{No} & \multicolumn{2}{c}{Yes} & \multicolumn{2}{c}{Total} \\
\hline
No & 3115 & (32.7\%) & 366 & (3.8\%) & 3481 & (36.6\%) \\ 
Yes & 4890 & (51.3\%) & 1152 & (12.1\%) & 6042 & (63.4\%) \\ 
\hline
Total & 8005 & (84.1\%) & 1518 & (15.9\%) & 9523 & \\
\end{tabular}
\end{centering}
\end{table}

Drilling further in the association between code coverage and crashes,
we examined the relationship between the covered methods, methods of 
unit-tested classes, and methods of the topmost stack frame.
Among 1\,166 methods associated with failures,
test code coverage and the existence of unit tests within the class
are related as depicted in Figure~\ref{fig:sharepie}
and summarized in Table~\ref{tab:ccvstest}.\footnote{Also
available in the ``Metrics'' sheet of the paper's replication package.}
The numbers we obtained indicate that code coverage on its own cannot
be used as a reliable indicator for determining the existence of
unit tests.
Consequently, we decided to consider as unit tested methods those
whose code is covered during testing {\em and} whose class has a
corresponding one with unit tests.

\subsection{Preliminary Qualitative Analysis} 
\label{sec:preliminaryqualitative}
We conducted a qualitative study of our data so as to gain a better insight 
of the crashes, the tests, and the code coverage results.

As expected, in the JaCoCo results found methods with code coverage
higher than 0 but no tests for their methods.
The reason for this is because JaCoCo 
in common with other code coverage tools shows which instructions, lines, or
branches of the code were (or were not) executed when running the tests.
This however does not mean that a given method was tested, because
its code might have been called by another method's test.
Below we outline specific cases of test coverage we related with
the existence of actual test code, as outlined in Section~\ref{sec:classes}.
 
\paragraph{1. Methods of classes with unit tests and zero code coverage}
There are methods with zero code coverage percentage
(according to the JaCoCo results)
that belong to a class that is associated with a test class found through
the heuristics outlined in Section~\ref{sec:classes}.
Since it does not make much sense to have unit tests
that are not executed,
we investigated this further to see why that happens.
The main reason seems to be wrong results from JaCoCo,
which skipped some tests that were not running due to configuration settings.
For example, most of the tests of the 
module \li{rt.equinox.p2} did not run successfully and so most of its classes and methods
got a zero coverage percentage, although they had unit tests.
Specifically, the class  \li{org.eclipse.equinox.internal.p2.metadata.repository.CompositeMetadataRepository}
is related to the unit-test class \li{org.eclipse.equinox.p2.tests.metadata.repository.CompositeMetadataRepositoryTest} which tests some of its methods,
but JaCoCo indicated zero coverage for its methods such as  the method  \li{addchild}.
Interestingly, this method also appeared on the topmost stack frame of 50 incident reports.

Another case occurs when the test class contains tests for a subset
of the class's methods and none of the other unit tests or unit-tested methods 
execute some methods.
This results in JaCoCo indicating no coverage for some methods.
Thus the presence of unit tests for a class is no guarantee for
unit tests for all the class's methods.
We found for example this to be the case in the method 
\li{doSetValue(Object source, Object value)} of the class 
\li{org.eclipse.core.internal.resources.Workspace}.
 
Despite its intuitive justification, this is not a common phenomenon:
only 336 methods of those listed in incident stack traces out a
total of 9\,523 belong to this category.

\paragraph{2. Methods of classes without unit tests and non-zero code coverage}
This category comprises
methods with a non-zero coverage percentage that do not belong to any of
the classes with unit test found through the procedure described
in section~\ref{sec:classes}.
This is an extension of the case where methods have their code covered,
even though they have no tests associated with them.
Unsurprisingly, we found 4\,890 methods out of 9\,523 belonging to this category,
because many methods delegate some work to others.

\paragraph{3. Methods of classes with unit tests and non-zero code coverage}
This category,
comprises methods that not only have a non-zero code coverage
percentage (according to JaCoCo), but also belong to a class with
unit-test code. There are 1\,152 methods belonging to this category 
(see Section~\ref{sec:quant}).
These two conditions provide the greatest assurance that a method is indeed
covered by a test.
In practice, we found two cases:
\begin{enumerate}
\item methods having non-zero code coverage and
a test class but no unit test for the specific method, and
\item methods having non-zero code coverage and
a test class with a unit test, which can fully or partially
cover the specific method. 
\end{enumerate}

An example of the first case can be found in method
\li{delete(int updateFlags, IProgressMonitor monitor)} of the
class \li{org.eclipse.core.internal.resources.Resource} 
and the corresponding unit test class \li{org.eclipse.core.tests.resources.ResourceTest}.
The method \li{delete} is executed by the method 
\li{ensureDoesNotExistInWorkspace} of the test class,
but there is no test for checking the specific method.

A representative example of the second case is the fully covered 
method \li{getAllSupertypes0}
of the class \li{org.eclipse.jdt.internal.core.hierarchy.TypeHierarchy}.
The corresponding test class
\li{org.eclipse.jdt.core.tests.model.TypeHierarchy} 
contains the method \li{testGetAllSupertypes}, which tests the method \li{getAllSupertypes0}.

An example of a partially covered method is the 
method \li{findContentViewerDescriptor}
of the class \li{org.eclipse.compare.internal.CompareUIPlugin} which is 
covered by 50\% on instruction level and 35\% by branch coverage.
The corresponding test class is the \li{org.eclipse.compare.tests.CompareUIPluginTest} and it  
contains the methods \li{testFindContentViewerDescriptor\_UnknownType}, 
\li{testFindContentViewerDescriptor\_TextType\_NotStreamAccessor}, 
\li{testFindContentViewerDescriptorForTextType\_StreamAccessor},
 which tests the method \li{findContentViewerDescriptor} providing 
 different inputs.
 
An excerpt of \li{findContentViewerDescriptor} in Listing~\ref{l:findContentViewerDescriptor} depicts multiple if statements.  According to JaCoCo, this method consists of 62 branches of which 
only the 23 are covered. On line 4 we can see for example the \li{TextType}
and on line 3 the \li{StreamAccessor} condition,
which seems to be related to test \li{testFindContentViewerDescriptorForTextType\_StreamAccessor}(Listing~\ref{l:findContentViewerDescriptorTests}).

\begin{lstlisting}[language=Java,
float,
caption={Code of partially covered method},
label={l:findContentViewerDescriptor},
numbers = left
]
public ViewerDescriptor[] findContentViewerDescriptor(Viewer oldViewer, Object in, CompareConfiguration cc) {
	Set result = new LinkedHashSet();
	if (in instanceof IStreamContentAccessor) {
		String type= ITypedElement.TEXT_TYPE;
		
		if (in instanceof ITypedElement) {
			ITypedElement tin= (ITypedElement) in;
				    
		    IContentType ct= getContentType(tin);
			if (ct != null) {
				initializeRegistries();
				List list = fContentViewers.searchAll(ct);
				if (list != null)
					result.addAll(list);
			}
		    
			String ty= tin.getType();
			if (ty != null)
				type= ty;
		}
\end{lstlisting}

\begin{lstlisting}[language=Java,
float,
caption={Test class of partially covered method},
label={l:findContentViewerDescriptorTests}
]
public void testFindContentViewer DescriptorForTextType_StreamAccessor() {
	CompareConfiguration cc = new CompareConfiguration();
	DiffNode in = new DiffNode(new TextTypedElementStreamAccessor(), new TextTypedElementStreamAccessor());
	ViewerDescriptor[] result = CompareUIPlugin.getDefault().
	findContentViewerDescriptor(null, in, cc);

	assertNotNull(result);
	assertEquals(1, result.length);
}

\end{lstlisting}

\paragraph{4. Non-faulty methods in incident stack traces}
We noticed that there are numerous methods that
appear in many incident stack traces,
although they have unit tests associated with them and appear to be
correct.
This raises the question of how could a method
appearing in so many incident reports
not have been noticed and fixed by the developers. 

One answer is that many of those methods are used for 
either debugging (such as
reporting exceptions, 
log specific messages,
handle assertions,
check if something exists or is null) 
or for triggering code
(such as \li{run()}, \li{invoke()}, \li{execute()})
and are therefore not directly associated a fault. 

The debugging methods are usually found at the first stack frame
of the stack trace, 
and the triggering methods within the top-6 and top-10 frames.
For example method 
\li{run(IWorkspaceRunnable action, IProgressMonitor monitor)} 
of class  \li{org.eclipse.core.internal.resources.Workspace} 
appeared in 4293 stack traces (such as \li{incident\_360023.json})
at the top-6 frames and method \li{checkExists} of class 
\li{org.eclipse.core.internal.resources.Resource} appeared in 2020 
stack traces (such as \li{incident\_1908559.json}) at the top frame.

\begin{lstlisting}[language=Java,
float,
caption={Faulty method's call chain code},
label={l:faulty}
]
public class SecurePreferences {
	[...]
	protected SecurePreferencesRoot getRoot() {
		if (root == null) {
			SecurePreferences result = this;
			while (result.parent() != null)
				result = result.parent();
			root = (SecurePreferencesRoot) result;
		}
		return root;
	}
	[...]
	public SecurePreferences parent() {
		checkRemoved();
		return parent;
	}
	[...]
	private void checkRemoved() {
		if (removed)
			throw new IllegalStateException(NLS.bind 
			(SecAuthMessages.removedNode, name));
	}
	[...]
}
\end{lstlisting}

\begin{lstlisting}[language=java,
float,
caption={Faulty method's call chain},
label={l:stack}
]
// incident_2029150.json
"stacktraces": [
   [
      {
        "cN": "org.eclipse.equinox.internal.security
        .storage.SecurePreferences",
        "fN": "SecurePreferences.java",
        "lN": 354,
        "mN": "checkRemoved"
      },
      {
        "cN": "org.eclipse.equinox.internal.security
        .storage.SecurePreferences",
        "fN": "SecurePreferences.java",
        "lN": 74,
        "mN": "parent"
      },
      {
        "cN": "org.eclipse.equinox.internal.security.storage
        .SecurePreferences",
        "fN": "SecurePreferences.java",
        "lN": 136,
        "mN": "getRoot"
      },
      [..]
      {
        "cN": "org.eclipse.core.internal.jobs.Worker",
        "fN": "Worker.java",
        "lN": 55,
        "mN": "run"
      }
   ]
]
\end{lstlisting}

In addition, there are methods further from the top stack frame
that just happen to be included in the stack trace
as part of a faulty method's call chain,
which also includes the faulty method that generated the stack trace.
For example, in Listing~\ref{l:stack}, which illustrated
the stack trace corresponding to the code in Listing~\ref{l:faulty},
the method \li{parent} only calls method \li{checkRemoved},
but stills appears in the incident stack trace,
although it is not the faulty method.

Method \li{checkRemoved} is at the top of the stack, because that
is where the application generated the stack trace. 
The \li{run()} method is at the bottom of the stack,
because this is how the program started. 
When the program started, the Java runtime executed the \li{run()} method. 
The \li{run()} method called \li{getRoot} and \li{getRoot} called \li{parent},
which called \li{checkRemoved}. 
Finally, \li{checkRemoved} threw \li{IllegalStateException},
which generated the stack trace.

\subsection{Statistical Analysis and Methods} 
\label{sec:stat}
A typical Java stack trace is a list of the method calls or stack frames 
that the application was in the middle of before an error or exception was thrown 
(or generated manually). A stack trace can range from a single stack frame 
(e.g stack trace of file \li{incident\_627736.json}) to 1024 frames 
(e.g stack trace of file \li{incident\_1655649.json}), with average length of 25 frames. 
The position of faulty methods in a stack trace can be found in one of 
the top-10 stack frames according to Schroter and 
his colleagues~\cite{schroter2010stack}, which mean that one of the last 10 methods 
that were called are likely to contain the defect. 

As described in Section~\ref{sec:preliminaryqualitative}, 
there are many methods appearing at top-6 and top-10 positions
(excluding the topmost position) that are not related to the crash.
Along with the fact that the exception occurred in the first stack frame, we consider the first (topmost) frame in the stack as the method that
caused the crash, and thus we decided to strictly define a method as associated
with the crash when it appears in the stack trace's top position. 

In addition, on our preliminary qualitative analysis,
we considered a method as unit-tested in one of the following cases:

\begin{enumerate}
\item
Method belongs to a unit-tested class but has no JaCoCo coverage data.
\item
Method does not belong to a unit-tested class but has JaCoCo coverage data.
\item
Method belongs to a unit-tested class {\em and} has JaCoCo coverage data.
\end{enumerate}

We decided to consider a method as unit tested 
by keeping only the methods that belong to 
a unit-tested class {\em and} have JaCoCo line code coverage
more than the median of the non-zero covered lines percentages,
namely 98.3\% .

Based on these two definitions we generate a $2 \times 2$ contingency table
containing the multivariate frequency distribution of the two variables:
tested, crashed.
We can then test for statistical significance
(deviation from the null hypothesis of RQ1) by applying Fisher's
exact test for count data~\cite{Fis22,Fis35},
which is available on the R statistical
analysis environment~\cite{IG96,R} as  \li{fisher.test}.
As we are only interested on whether testing is associated with
fewer crashes (and not the reverse of whether fewer crashes are
associated with testing) we test the alternative hypothesis
in the {\em less} direction.

\section{Results} 
\label{sec:results}
Here we answer our two research questions by means of statistical (RQ1)
and qualitative (RQ2) analysis.
\subsection{Statistical Analysis} 
\begin{figure}
{
\includegraphics[width=0.55\columnwidth]{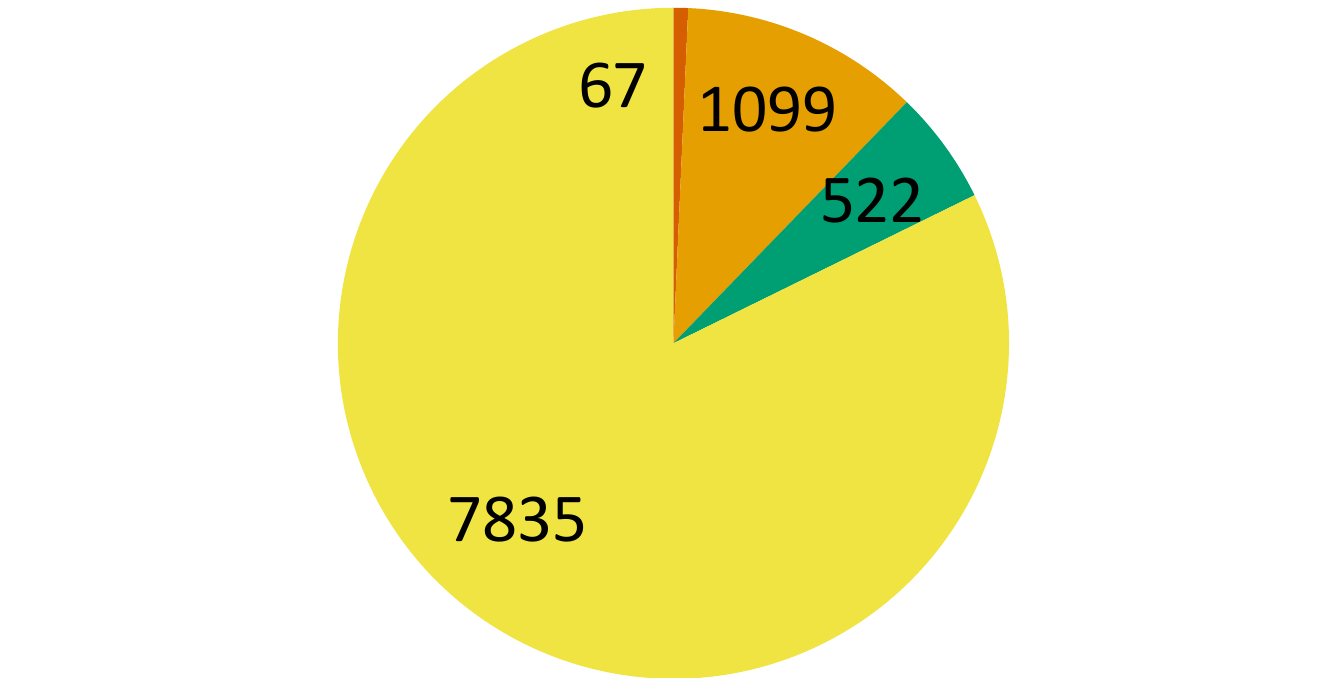}
\includegraphics[width=0.42\columnwidth]{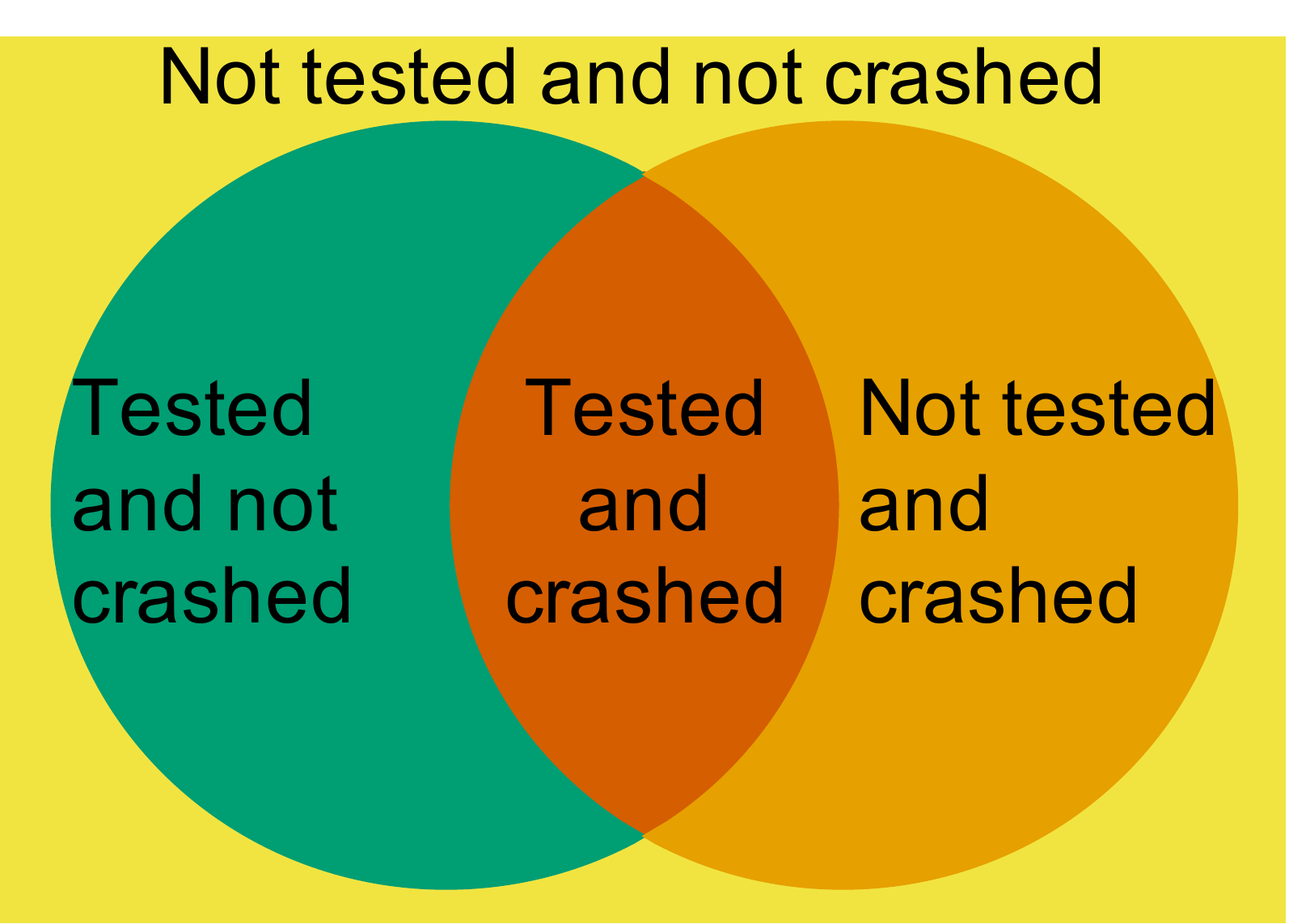}
}
\caption{Relationship between strictly tested and strictly crashed methods.
The pie areas correspond to the colored areas of the Venn diagram.
}
\label{fig:sharepie2}
\end{figure}

\begin{table}
\caption{Crashes of Tested Methods}
\label{tab:results}
\begin{centering}
\begin{tabular}{lrrrrrr}
\hline
 & \multicolumn{4}{c}{Unit tested} & & \\
Crashed & \multicolumn{2}{c}{No} & \multicolumn{2}{c}{Yes} & \multicolumn{2}{c}{Total} \\
\hline
No & 7835 & (82.3\%) & 522 & (5.5\%) & 8357 & (87.8\%) \\ 
Yes & 1099 & (11.5\%) & 67 & (0.7\%) & 1166 & (12.2\%) \\ 
\hline
Total & 8934 & (93.8\%) & 589 & (6.2\%) & 9523 & \\
\end{tabular}
\end{centering}
\end{table}

To answer RQ1 regarding the association between unit tests and
crashes we
classified the 9\,523 methods as unit tested and crashed according
to the criteria we specified in Section~\ref{sec:stat}.
This resulted in their categorization depicted
in Figure~\ref{fig:sharepie2} and and summarized in Table~\ref{tab:results}.

Our question is whether testing a piece of code is associated
with a lower chance of it crashing.
Applying Fisher's exact test for count data,
results in a $p$-value of 0.278 and an
odds ratio based on the conditional maximum likelihood estimate
of  0.915 in a 95\% confidence interval of 0--1.146.
Consequently, we cannot reject the null hypothesis,
and conclude that
{\bf our data set does not provide statistical evidence
supporting the hypothesis that
the presence of unit tests is associated with fewer crash incidents}.

\subsection{Qualitative Analysis} 
\label{sec:qualitative}

To answer RQ2 on why do unit-tested methods still fail, 
we analyzed the 67 methods that are strictly unit-tested and crashed.
This may sound like a small number,
but those methods appeared in 10\,608 stack traces.

By examining the stack traces and their relevant methods in the Eclipse
source code, 
we classified crashes of unit tested methods into three categories.

\paragraph{1. The method contains a developer-introduced fault}

These faults stem from programmer errors, such as
algorithmic, logic, ordering, dependency, or consistency
errors~\cite{ko2005framework}.
They mainly involve code parts that are missing error-handling mechanisms
for code that can potentially throw exceptions,
thus causing the application to crash with an uncaught exception error.

An example of a method belonging to this category, is
\li{getPath} of class 
\li{org.eclipse.jdt.internal.core.search.JavaSearchScope.java}~(Listing~\ref{l:getpath}).
This was found to be the topmost method 
in 12 stack traces, such as \li{incident\_69854.json}.
The stack trace indicated that the crash occurred 
on line 2  which makes sense, because a \li{NullPointerException} can
be thrown at that point if \li{element} is {\em null}.

This method belongs to the category 3.1 we presented in 
Section~\ref{sec:preliminaryqualitative}.
It has been called by multiple other tested 
methods providing different element input, without testing it with a {\em null}
argument.
If developers had written a test that specifically checked this method, 
they might have covered this case.

\begin{lstlisting}[language=Java,
float,
caption={Example of a faulty method caused by a developer error regarding an invalid argument},
label={l:getpath},
numbers=left
]
private IPath getPath(IJavaElement element, boolean relativeToRoot) {
	switch (element.getElementType()) {
		case IJavaElement.JAVA_MODEL:
			return Path.EMPTY;
		case IJavaElement.JAVA_PROJECT:
			return element.getPath();
		// [...]
		default:
			return getPath(element.getParent(), relativeToRoot);
	}
}
\end{lstlisting}

Another example is the method \li{consumeEmptyStatement}
~(Listing~\ref{l:consumeEmptyStatement}).
This was found to be the topmost method 
in 11 stack traces such as \li{incident\_1324714.json}) of class 
\li{org.eclipse.jdt.internal.compiler.parser.Parser}.
It seems to cause an \li{IndexOutOfBoundsException} on line 3.

\begin{lstlisting}[language=Java,
float,
caption={Example of a faulty method caused by a developer error regarding array indexing},
label={l:consumeEmptyStatement},
numbers=left
]
protected void consumeEmptyStatement() {
 char[] source = this.scanner.source;
 if (source[this.endStatementPosition] == ';') {
   pushOnAstStack(new EmptyStatement (this.endStatementPosition, this.endStatementPosition));
 } else {
\end{lstlisting}

\paragraph{2. The method intentionally raises an exception}
There are methods that intentionally lead to crashes due to internal errors,
wrong configuration settings, or unanticipated user behavior,
rather than faults introduced through a developer oversight.
Developers understood that these failures could potentially happen 
under unforeseen circumstances or in ways that could not be appropriately
handled.
As a backstop measure they intentionally throw exceptions 
with appropriate messages in order to log the failure and collect
data that might help them to correct it in the future.

\begin{lstlisting}[language=Java,
float,
caption={Example method of an internal error},
label={l:checkremoved}
]
private void checkRemoved() {
	if (removed)
		throw new IllegalStateException(NLS.bind (SecAuthMessages.
		removedNode, name));
}
\end{lstlisting}
We found for example this to be the case in  stack trace \li{incident\_2029150.json} in which the first frame contains the method  \li{checkRemoved} ~(Figure~\ref{l:checkremoved})
of class \li{org.eclipse.equinox.internal.security.storage.SecurePreferences}, which throws an \li{IllegalStateException}
and logs the message {\em ``Preference node '\$node' " has been removed''}.

\paragraph{3. The method is not faulty}
\begin{lstlisting}[language=Java,
float,
caption={Example of a non-faulty method},
label={l:fail}
]
private void fail(String message) throws TemplateException {
	fErrorMessage= message;
	throw new TemplateException(message);
}
\end{lstlisting}
There are methods in the topmost stack position that
simply report a failure associated with a fault in another method.
These methods are the non-faulty (debugging) methods we described
in category 4 of Section~\ref{sec:preliminaryqualitative}.
An example of such a case is method 
\li{fail} of the class \li{org.eclipse.jface.text.templates.TemplateTranslator}~(Figure~\ref{l:fail}).

Having analyzed the crashes, we worked on understanding why 
those crashes occurred while there was (apparently) unit tested code. 
Unit testing would not help alleviate cases 2 and 3,
and therefore we did not investigate further. 
On the other hand, methods belonging to the first case are
much more interesting, so we dug deeper to understand
the types of faults, failures, and their relationship to unit testing,
and categorized them into the following areas.

\paragraph{1.a Method is not called by the class's tests}
The method's test class does not call the specific method
in any of the tests.
The method may have been incidentally called by tests of other classes.
We found for example this to be the case in method  \li{resetProcessChangeState} of class 
\li{org.eclipse.text.undo.DocumentUndoManager}.
 
\paragraph{1.b Method is not tested by the class's tests}
The method's test class calls this method to setup or validate
other tests, but does not explicitly test the given method.
We found for example this to be the case in 
method \li{getPath}~(Listing~\ref{l:getpath}) we previously saw on this section on 
paragraph 1.
Similarly, there are many such tests in 
test class \li{org.eclipse.jdt.core.tests.compiler.parser.ParserTest} 
or \li{org.eclipse.core.tests.resources.ResourceTest}.

\paragraph{1.c Specific case is not tested}
The method has a unit test, but some specific cases are not tested.
Ideally, all cases should be tested to ensure that the discrete unit
of functionality performs as specified under all circumstances.
An example of this case is the partially (90\%) covered  method \li{iterate}
of the class \li{org.eclipse.core.internal.watson.ElementTreeIterator} with unit test 
class \li{org.eclipse.core.tests.internal.watson.ElementTreeIteratorTest}. 
This method causes an \li{IndexOutofbounds} Exception,
but the test method does not test this case.

\section{Discussion} 
\label{sec:discussion}

In isolation and at first glance, the results we obtained are startling.
It seems that unit tested code is not significantly less likely to be
involved in crashes.
However, one should keep in mind that absence of evidence is not
evidence of absence.
We have definitely {\em not} shown that unit tests {\em fail} to reduce crashes.

Regarding the result,
we should remember that our data come from a production-quality
widely used version of Eclipse.
It is possible and quite likely that the numerous faults resulting
in failures were found through the unit tests we tallied in earlier
development, alpha-testing, beta-testing, and production releases.
As a result, the tests served their purpose by the time the particular
version got released, eliminating faults whose failures do not
appear in our data set.

Building on this, we must appreciate that
not all methods are unit tested and not all methods are unit tested
with the same thoroughness.
Figure~\ref{fig:sharepie} shows that fewer than half of the methods
and lines are unit tested.
Furthermore, Figure~\ref{fig:ccoverall} shows that code coverage within
a method's body also varies a lot.
This may mean that developers selectively apply unit testing
mostly in areas of the code where they believe it is required.

Consequently,
an explanation for our results can be that unit tests are preferentially added
in complex and fault-prone code in order to weed out implementation bugs.
Due to its complexity, such code is likely to contain further
undetected faults,
which are are in turn likely to be involved in field failures
manifesting themselves as reported crashes.

One may still wonder how can unit-tested methods with a 100\% code coverage
be involved in crashes.
Apart from the reasons we identified in Section~\ref{sec:qualitative},
one must appreciate that test coverage is a complex and elusive concept.
Test coverage metrics involve
statements,
decision-to-decision paths (predicate outcomes),
predicate-to-predicate outcomes,
loops,
dependent path pairs,
multiple conditions,
loop repetitions, and
execution paths~\cite[pp. 142--145]{Jor02},~\cite{BMS98}.
In contrast, JaCoCo analyses coverage at the level of
instructions, lines, and branches.
While this functionality is impressive by industry standards,
predicate outcome coverage can catch only about 85\% of revealed
faults~\cite[p. 143]{Jor02}.
It is therefore not surprising that failures still occur in unit tested code.

An important factor associated with our results is that failures
manifested themselves exclusively through exceptions.
Given that we examined failure incidents through Java stack traces,
the fault reporting mechanism is unhandled Java exceptions.
By the definition of an unhandled exception stack trace,
all methods appearing in our data set passed an exception through
them without handling it internally.
This is important, for two reasons.
First, unit tests rarely examine a method's exception processing;
they typically do so only when the method under test
is explicitly raising or handling exceptions.
Second, most test coverage analysis tools fail to report coverage of
exception handling, which offers an additional, inconspicuous, branching
path.

It would be imprudent to use our findings as an excuse to avoid unit testing.
Instead, practitioners should note that unit testing on its own is not
enough to guarantee a high level of software reliability.
In addition, tool builders can improve test coverage analysis systems
to examine and report exception handling.
Finally, researchers can further build on our results to recommend
efficient testing methods that can catch the faults that appeared in
unit tested code and test coverage analysis processes to pinpoint
corresponding risks.

\section{Threats to Validity} 
\label{sec:threats}

Regarding external validity,
the generalizability of our findings is threatened by our choice of
the analyzed project. Although Eclipse is a very large and
sophisticated project, serving many different application areas,
we cannot claim that our choice represents adequately all software
development. For example, our findings may not be applicable to
small software projects,
projects in other application domains,
software written in other programming languages, or
multi-language projects.
Finally, we cannot exclude the possibility that the selection of
a specific Eclipse product and release may have biased our results.
If anything, we believe additional research should look at
failures associated with less mature releases.

Regarding internal validity we see four potential problems.
First, the code coverage metrics we employed have room for improvement,
by incorporating e.g. branch coverage or mutation testing data.
Second, employing JaCoCo on an old release which may have some
deprecated code and archived repos, caused some unit test failures,
resulting in a lower code coverage.
Third, we excluded
from the JaCoCo report non-Java code that is processor architecture specific
(e.g the \li{org.eclipse.core.filesystem.linux.x86} bundle).
Fourth, noise in some meaningless stack frames appearing in our
stack trace dataset may have biased the results.

\section{Related Work} 
\label{sec:related}
Among past studies researching the relationship between
unit test coverage and software defects, the most related to our work
are the ones that examine actual software faults.
Surprisingly, these studies do not reach a widespread agreement
when it comes to the relationship between the two.
More specifically existing findings diverge regarding the hypothesis
that a high test coverage leads to fewer defects.
Mockus et al. \cite{MND09}, who studied two different industrial
software products, agreed with the hypothesis and concluded that code
coverage has a negative correlation with the number of defects.
On the other hand, Gren and Antinyan's work~\cite{gren2017relation}
suggests that unit testing coverage is not related to fewer defects and there is no strong relationship between unit testing and code quality. A more recent study by the same primary author~\cite{8354427},
investigated an industrial software product,
and also found a negligible decrease in defects when coverage increases,
concluding that test unit coverage is not a useful metric for test effectiveness. 

Furthermore, in a study of seven Java open source projects,
Petric et al. found that the majority of methods with defects had
not been covered by unit tests~\cite{petric2018effectively},
deducing that the absence of unit tests is risky and can lead to failures.
On the other hand, Kochhar et al. in another study of one hundred Java projects,
did not find a significant correlation between code coverage and
defects~\cite{KLLN17}.

The above mentioned studies cover only fixed faults.
In our research, we work with stack traces, which enable us to
analyze field-reported failures associated with crashes.
The associated faults include those that have not been fixed,
but exclude other faults that are not associated with crashes,
such as divergence from the expected functionality or program freezes.
Furthermore, through the crash reports we were unable to know
the faulty method associated with the crash.
However, by placing our matched crash methods in three groups
according to their respective position in the stack trace
(in the very first stack frame, within the top-6 and the top-10 stack frames)
we could obtain useful bounds backed by empirical evidence~\cite{schroter2010stack}
regarding the coverage of methods that were likely to be defective.

Considerable research
associating testing with defects
has been performed on the relationship between
test-driven development and software defects.
Test-driven development (TDD) is centered around rapid iterations
of writing test cases for specifications and then the corresponding
code~\cite{Bec03}.
As a practice it obviously entails more than implementing
unit tests, but absence of evidence of TDD benefits should also
translate to corresponding absence of benefits through simple unit testing,
though the benefits of TDD will not necessarily translate into benefits
of unit testing.
In a review of the industry’s and academia’s empirical studies, 
Mäkinen and Münch~\cite{MM14} found that TDD
has positive effects in the reduction of defects a result also mirrored
in an earlier meta-analysis~\cite{rafique2012effects}
and a contemporary viewpoint~\cite{munir2014considering}.
In industry, an IBM case study found that test-driven development
led to 40\% fewer defects~\cite{williams2003test}.
In academia, classroom experiment results showed
that students produce code with 45\% fewer defects using TDD~\cite{edwards2003using}.
On the other hand, experimental results by Wikerson and Mercer failed to
show significant positive effects~\cite{wilkerson2011comparing}.

The study by Jia and Harman~\cite{jia2010analysis} shows clear evidence that mutation testing has gained a lot popularity during the past years.
The majority of researchers concluded that high mutation score improves fault detection \cite{papadakis2018mutation}. Furthermore, mutation testing can reveal additional defect cases beyond real faults \cite{andrews2005mutation}. However, mutants can only be considered substitute of real faults under specific circumstances \cite{just2014mutants}.

Apart from Schroter and his colleagues~\cite{schroter2010stack},
a number of researchers have studied the Eclipse IDE
and most of them have focused on predicting defects.
Most notably, Zimmermann and his colleagues provided 
a dataset mapping defects from the Eclipse database to
specific source code locations annotated with common complexity
metrics~\cite{zimmermann2007predicting}, while
Zhang~\cite{zhang2009investigation} based on Eclipse data,
yet again, suggested lines of code as a simple but good predictor of defects.

\section{Conclusions} 
\label{sec:conclusions}
Software testing contributes to code quality assurance and helps developers detect and correct
program defects and prevent failures.
Being an important and expensive software process activity it has to be efficient.
In our empirical study on the Eclipse project
we used the JaCoCo tool and a class source code matching procedure
to measure the test coverage, and
we analyzed field failure stack traces to assess the effectiveness of testing.
Our results indicate that unit testing on its own may not be a sufficient
method for preventing program failures.
Many methods that were covered by unit tests were involved in crashes,
which may mean that the corresponding unit tests were not sufficient for uncovering
the corresponding faults.
However, it is worth keeping in mind that failures manifested themselves
through exceptions whose branch coverage JaCoCo is not reporting.
Research building on ours can profitably study
the faults that led to the failures we examined in order to propose
how unit testing can be improved to uncover them, and
how test coverage analysis can be extended to suggest these tests.

\begin{acks}
We thank Philippe Krief and Boris Baldassari for their invaluable help
regarding the Eclipse incident data set.
Panos Louridas provided insightful comments on an earlier version of this
manuscript.
Dimitris Karlis expertly advised us on the employed statistical methods.
This work has been partially funded by:
the CROSSMINER project,
which has received funding from the
\grantsponsor{H2020}{European Union Horizon 2020}{https://ec.europa.eu/programmes/horizon2020}
research and innovation programme under grant agreement No~\grantnum{H2020}{732223};
the FASTEN project,
which has received funding from the
\grantsponsor{H2020}{European Union Horizon 2020}{https://ec.europa.eu/programmes/horizon2020}
research and innovation programme under grant agreement No~\grantnum{H2020}{82532};
the \grantsponsor{GSRT}{GSRT}{http://www.gsrt.gr/}
2018 Research Support grant~\grantnum{GSRT}{11312201}; and
the \grantsponsor{RCAUEB}{Athens University of Economics and Business Research Centre}{https://www.aueb.gr}
Original Scientific Publications 2019 grant~\grantnum{RCAUEB}{EP-3074-01}.
\end{acks}

\bibliographystyle{ACM-Reference-Format}
\bibliography{crash-unit}
\end{document}